\documentclass[aip, amsmath,amssymb,reprint]{revtex4-1}

\usepackage{graphicx}
\usepackage{epstopdf}
\usepackage{dcolumn}
\usepackage{bm}
\usepackage{amssymb}
\usepackage{setspace}
\usepackage{dcolumn}
\usepackage{bm}

\usepackage{float}
\usepackage{array}
\usepackage{epsfig}

\begin{document}

\preprint{AIP/123-QED}

\title[]{Piezoelectric Tunable Microwave Superconducting Cavity}

\author{N. C. Carvalho}
\email[]{natalia.docarmocarvalho@research.uwa.edu.au}
\affiliation{School of Physics, The University of Western Australia, 35 Stirling Hwy, 6009 Crawley, Western Australia.}
\affiliation{ARC Centre of Excellence for Engineered Quantum Systems (EQuS), 35 Stirling Hwy, 6009 Crawley, Western Australia.}

\author{Y. Fan}
\affiliation{School of Physics, The University of Western Australia, 35 Stirling Hwy, 6009 Crawley, Western Australia.}

\author{M. E. Tobar}
\affiliation{School of Physics, The University of Western Australia, 35 Stirling Hwy, 6009 Crawley, Western Australia.}
\affiliation{ARC Centre of Excellence for Engineered Quantum Systems (EQuS), 35 Stirling Hwy, 6009 Crawley, Western Australia.}%

\date{\today}

\begin{abstract}
In the context of engineered quantum systems, there is a demand for superconducting tunable devices able to operate with high Q-factors at power levels equivalent to only a few photons. In this work, we developed a 3D microwave reentrant cavity with such characteristics ready to provide a very fine-tuning of a high-Q resonant mode over a large dynamic range. This system has an electronic tuning mechanism based on a mechanically amplified piezoelectric actuator, which controls the resonator dominant mode frequency by changing the cavity narrow gap by very small displacements. Experiments were conducted at room and dilution refrigerator temperatures showing a large dynamic range up to 4 GHz and 1 GHz, respectively, and were compared to a FEM model simulated data. At elevated microwave power input, nonlinear thermal effects were observed to destroy the superconductivity of the cavity due to the large electric fields generated in the small gap of the reentrant cavity.
\end{abstract}

\maketitle

\section{Introduction}

A microwave reentrant cavity consists of a cylindrical shaped structure with a narrow gap between a reentrant post and the cavity lid. Such a device presents an interesting electromagnetic field configuration with an azimuthal magnetic component around the post and an intense axial electric field in the gap spacing, which makes its dominant resonance frequency very sensitive to the gap dimensions. Electrical quality factors (Q-factors) up to 10$^8$ have already been reported when operated in the superconducting regime and undergone an intense cleaning process \cite{bassan2008}. Without such a rigorous cleaning Q-factors above $10^5$ are regularly achieved.

In fact, microwave reentrant cavities have being investigated over the past 50 years \cite{fujisawa1958, hansen1939} and repeatedly employed as transducers in resonant-mass gravitational wave detectors \cite{linthorne1992, barroso2004}, finding also applicability in different research areas, such as solid state microwave oscillators \cite{williamson1976}, particle accelerators \cite{grimm2005}, dielectric characterization \cite{baker1998}, electron spin resonance spectroscopy \cite{giordano1983} and test of fundamental physics \cite{mcallister2015}. 

Most of the aforementioned applications would benefit from a microwave cavity with a tunable superconducting mode, fine tunability and large dynamic range. For this reason, tunable cavities coupled to piezoelectric actuators have been developed for decades \cite{klotz1969, buscher1978}. More recently, a system was presented that uses the central post of a reentrant cavity as a movable structure and was able to offer fine electronic tuning over a range of up to 139 MHz at cryogenic temperatures, but with low Q-factors \cite{carvalho2014}. 

The new tuning mechanism developed in this work offers an improved dynamic range due to the mechanically Amplified Piezoelectric Actuator (APA) manufactured by Cedrat Technologies SA. We also gain a substantial increase in resonator Q-factor by combining a superconducting cavity with a different design, where a movable cavity lid with a choke mechanism was used to control the resonance frequency. This allowed the post to be fixed, reducing extra resistive losses associated with the post not being in proper electrical contact with the body of the resonant cavity.

The system is able to tune the cavity resonance frequency over ranges of several hundreds of MHz at 10 mK with Q-factors higher than 10$^5$. Moreover, this system does not degrade with only a few photons on average within the cavity, a condition required for the operation of hybrid quantum systems and Quantum Electrodynamics experiments, which necessarily couple the resonator to superconducting quantum devices such a qubit or a SQUID. 

In this context, a 3D microwave cavity would offer advantages over on chip resonators, being able to supply a strong and localized electromagnetic field with a low loss microwave mode. Moreover, to control the interaction between the cavity and the quantum device usually a magnetic field will be applied. However, this ability is limited by the critical field of the superconducting electronics, thus an electronic way to tune a resonator over a large dynamic range and under cryogenic conditions is a valuable alternative, alleviating the necessity of large DC magnetic fields.  

Aside from the tunability properties of the system, microwave nonlinear effects due superconducting-to-normal-state transition were observed to affect the cavity resonance. Such behavior can be associated to hot surface regions formation when the resonator is subject to high microwave power driving.

\section{Experimental}

\begin{figure}[h!]
\includegraphics[width=0.48\textwidth]{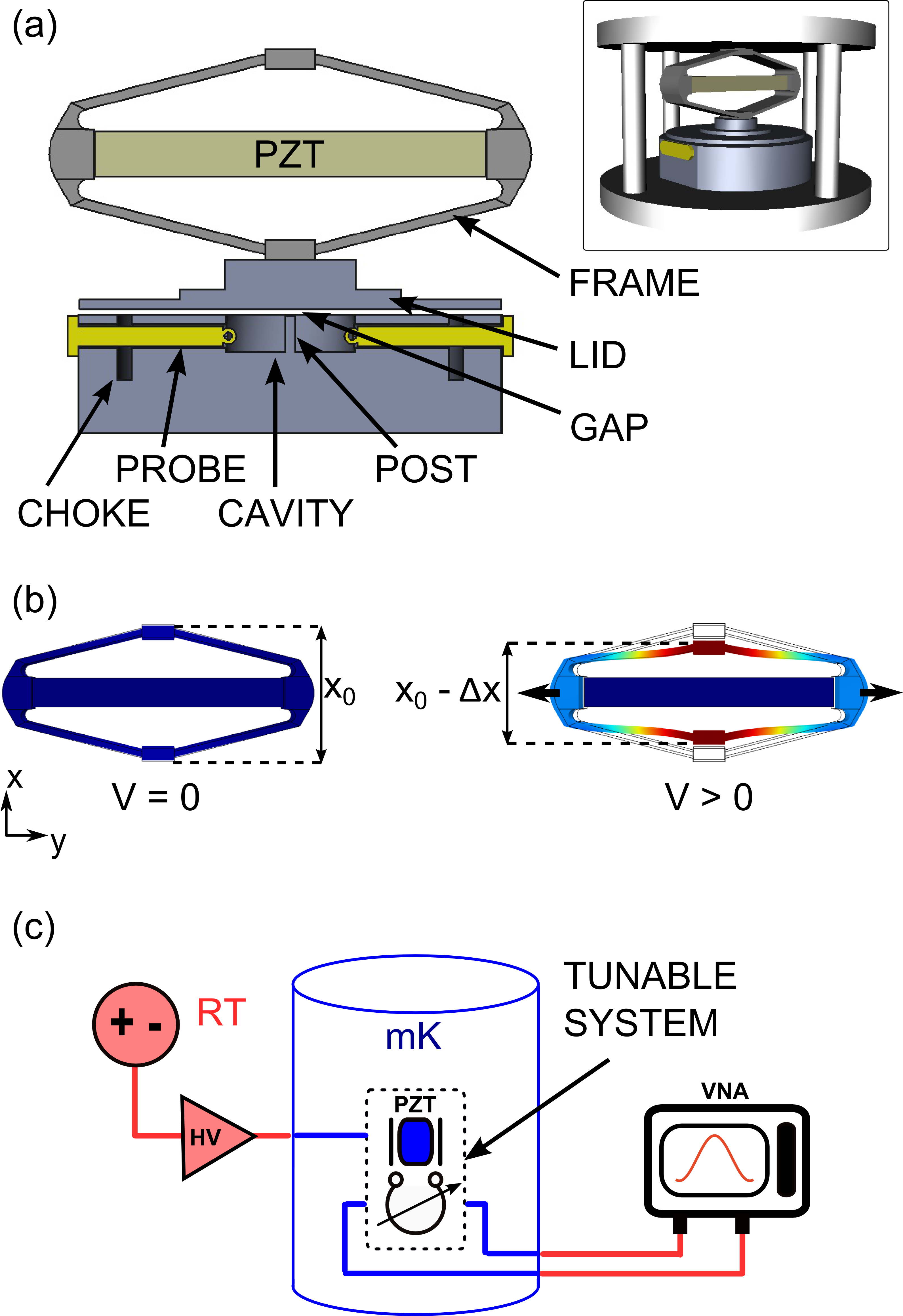}
\caption{\label{fig1}(a) Schematic of the microwave tunable cavity. Inset: three-dimensional illustration of the system. (b) APA frame deformation when a DC voltage is applied to the device. (c) Schematic diagram of the experiment setup for cryogenic measurements (HV: high voltage power amplifier; VNA: Vector Network Analyzer; PZT: piezoelectric actuator).}
\end{figure}

The detailed tunable system design is shown in Fig. \ref{fig1} (a). The cylindrical reentrant cavity is 3.7 mm high and its radius is equal to 6.4 mm with a central post whose diameter is 1 mm. The cavity is made of niobium and is located at the bottom of the system, where two magnetic probes can be coupled to the cavity field. Around the cavity is the choke: a cut ring with inner radius equal to one quarter of the resonant mode wavelength and whose function is to reduce the energy leakage between the cavity top and bottom.

The cavity lid is movable and not connected to the bottom structure, hence in order to tune the gap spacing between the post and the top wall the APA is attached to the lid. Then, by exciting the piezoelectric material through a difference of potential it has its initial length increased in the horizontal direction deforming the APA frame as in Fig. \ref{fig1} (b). The frame structure supplies mechanical amplification to the piezoelectric displacement and changes the initial frame height from $x_0$ to  $x_0 - \Delta x$, where $x_0$ is the initial vertical dimension. Therefore, because the top of the frame is held by a fixed support, the cavity lid is pulled up by $\Delta x$ and the displacement is translated into frequency tuning.  

The electronic tuning mechanism is shown in Fig. \ref{fig1} (c), where the diagram presents the tunable system located in the mixing chamber of the dilution refrigerator. Very thin power cables connect the APA to a high voltage power supply outside the refrigerator, providing up to 150 V, which is the maximum DC voltage supported by the actuator. 

The tunable cavity was also designed to allow initial tuning of the gap, which can be changed by adding or removing spacers between the APA frame and its holder. This enables the setting of an initial frequency between 6 GHz and 9 GHz, but requires the tunable system to be taken out of the refrigerator, as it can only be done manually. In summary, the initial frequency is set initially at room temperature and additional finer tuning with a dynamic range of up to a GHz can be provided by the electronic mechanism when the cavity is already cold. 

\section{Results and discussion}

\begin{figure}
\includegraphics[width=0.42\textwidth]{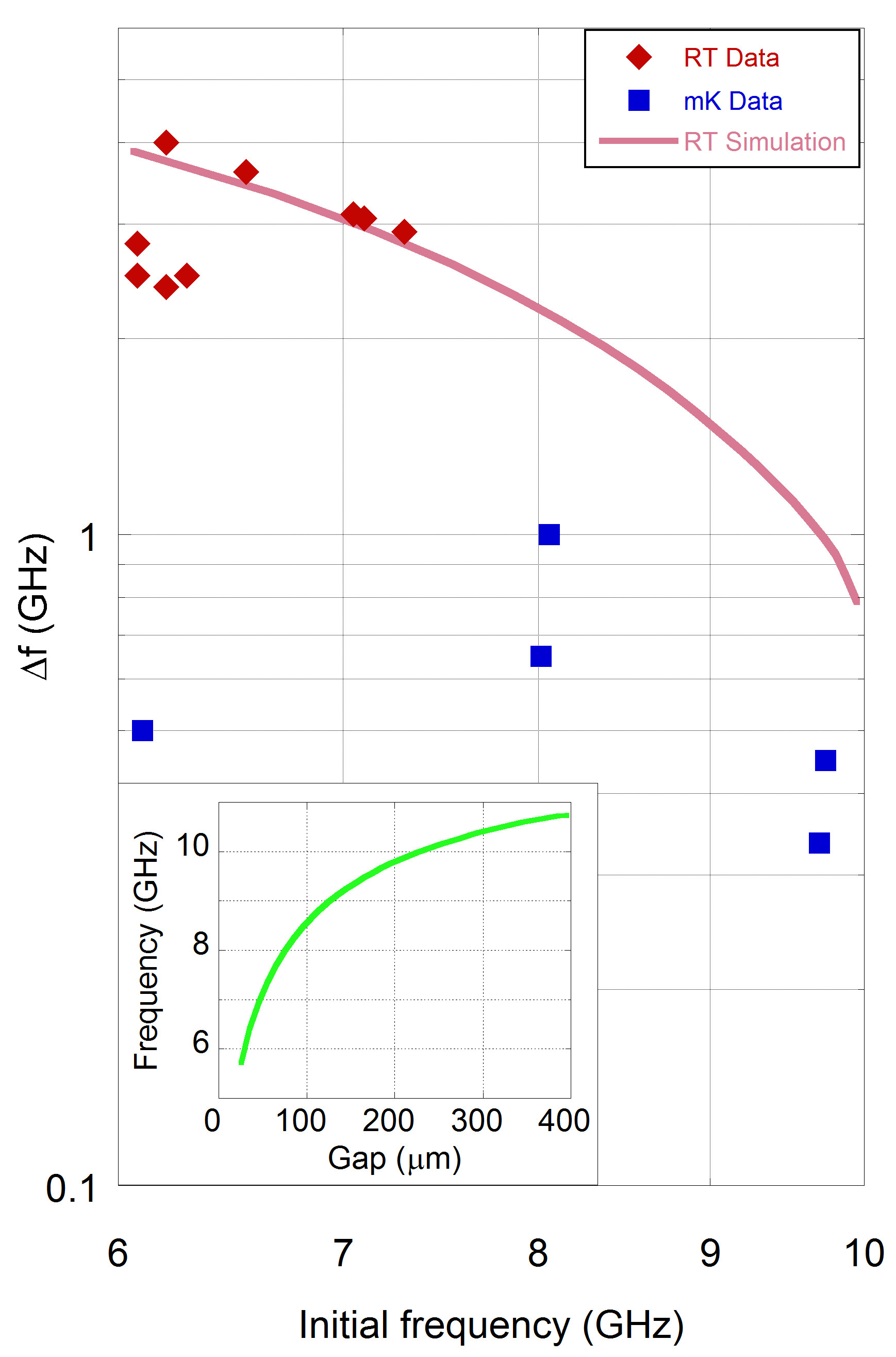}
\caption{\label{fig2}Measured dynamic range of the frequency tuning as a function of the mechanically set initial frequency for room temperature and cryogenic measurements. Inset: cavity displacement sensitivity simulated by the FEM model.}
\end{figure}

Fig. \ref{fig2} shows the measured frequency tuning range ($\Delta f$) for room temperature (RT) and cryogenic (mK) experiments as a function of the initially set resonance frequency. In this plot  $\Delta f$ is the frequency change only due the electronic tuning when the maximum voltage is applied to the APA. The results can be compared to simulation calculated from Eq. (\ref{Eq1}), where $\Delta f_{sim}$ is the simulated tuning range, $\Delta x_0$ is the APA frame vertical displacement after 150 V was applied to the actuator and $(\Delta f/ \Delta d)_{fem}$ is the cavity displacement sensitivity:

\begin{eqnarray}
\Delta f_{sim} = \Delta x_0 \times \left(\frac{\Delta f}{\Delta d}\right)_{fem}.
\label{Eq1}
\end{eqnarray}

A Finite Element Method (FEM) model was used to simulate the relationship between the cavity frequency (\textit{f}) and the gap spacing (\textit{d}). The curve used to calculate $(\Delta f/ \Delta d)_{fem}$ is presented in Fig. \ref{fig2} inset. The $\Delta x_0$ was estimated by neglecting the cavity lid weight and adopting the minimum possible frame vertical displacement at 150 V, which according to the APA's manufacturer is 189 $\mu$m. 

Experimental data were taken at room temperature and 10 mK. The former demonstrates reasonable consistency with the simulations, with many of the points agreeing very well with the simulated data. A few measurements close to 6 GHz exhibit a small decrease in the tuning range in comparison to the simulations. This is attributed to unwanted deformation of APA frame during the assembling process. Deformations on cooling to cryogenic temperatures are harder to control, and hence the tuning range is harder to predict and exhibits some scatter. Previous results have shown the dynamic range of a piezoelectric actuator when cooled to cryogenic temperatures is reduced by about $82\%$ \cite{carvalho2014}, in this case the reduction in tuning range compared to the simulated room temperature performance is 66\% on average, showing a less significant reduction in sensitivity compared to the previous results.

\begin{figure}
\includegraphics[width=0.48\textwidth]{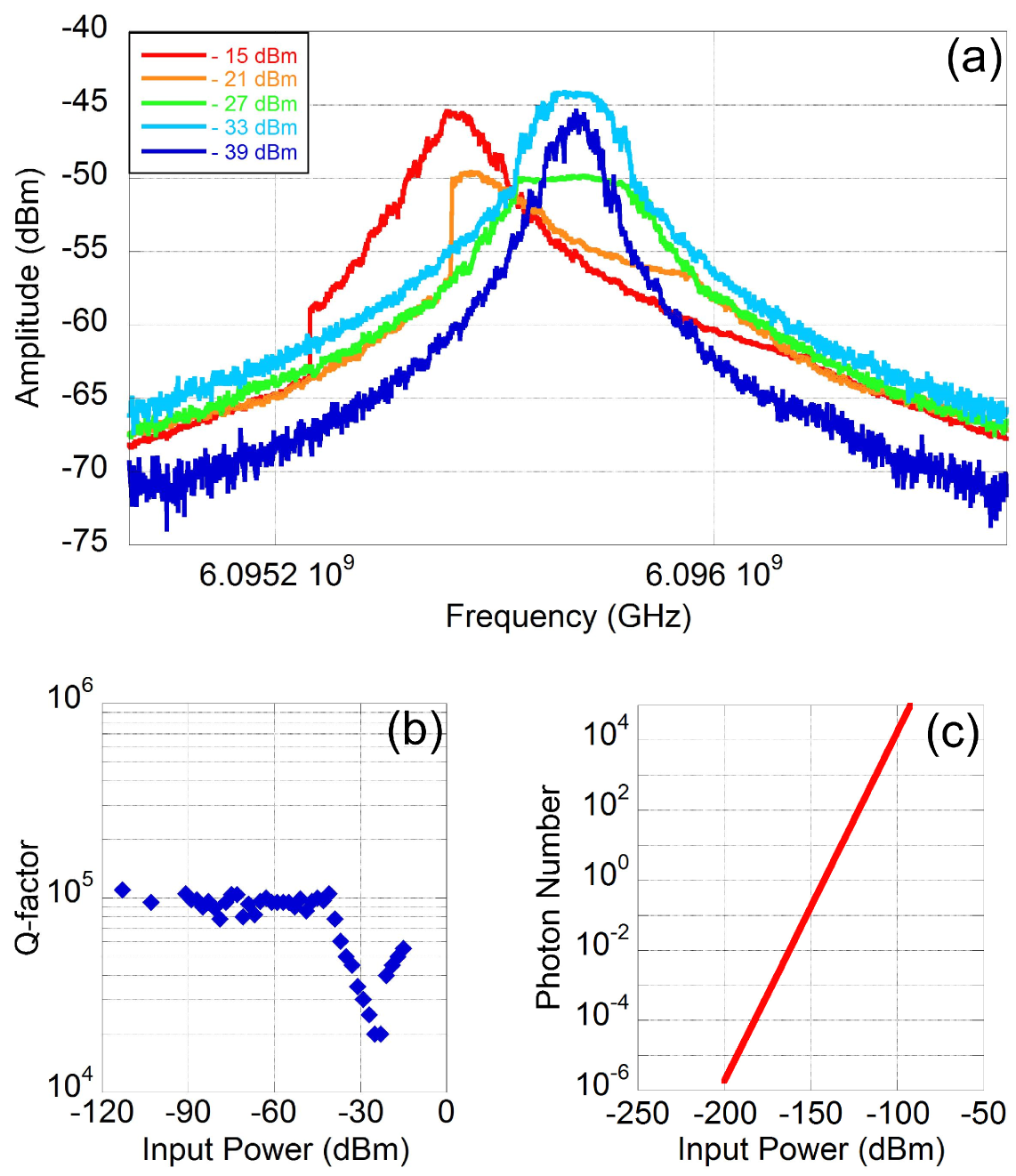}
\caption{\label{fig3}(a) Measured nonlinear resonance response due superconducting-to-normal-state transition. (b) Measured resonator Q-factor for different cavity input power. (c) Calculated cavity photon number for different input power.}
\end{figure}

The best tuning performance of the cryogenic trials shows that the cavity resonance frequency can be tuned over a range as large as 1 GHz. The aim, however, is to achieve a large tuning range combined to a high electrical Q-factor. From this perspective the best measured Q-factor, equal to 10$^5$, occurs when the initial frequency is 6.1 GHz, where the tuning range is about 500 MHz . Such result can be easily explained by leakage through the gap between cavity top and bottom. The choke was designed to minimize this effect; however, it is less efficient as the gap is increased. 

Nonlinear effects were also observed when the resonator was set to 6.1 GHz  and was operated at relatively high microwave powers. Fig. \ref{fig3} (a) has the transmission measurement of the cavity resonance for different input power showing a clear deformation of the resonance shape with increasing power. This effect besides distorting the Lorentzian curve, also shifted the resonance peak to lower frequencies.

Such behavior is often associated to the breakdown of the superconductivity regime on the resonator surface. The heating caused by elevated power intensity can increase the material surface resistance destroying the superconductivity in some hot regions and thus leading to the normal-state transition. As a consequence, this effect will degrade the resonator Q-factor and may change its resonance frequency \cite{anlage1999, kurter2011}. The narrow gap spacing of the reentrant cavity is the most likely region to become non-superconducting due to the intense local electric field, which in turn will cause the temperature to rise, explaining the flattening and broadening of the resonance peak when the cavity is overdriven by high power.

Fig. \ref{fig3} (b) shows how the effect of hot regimes influences the resonator Q-factor. From -5 to -15 dBm the resonator is  
predominantly in the normal regime, but superconducting regions begin to appear leading to a nonlinear response of the cavity resonance and, therefore, broadening the transmission spectrum as these regions spread over the resonator surface. The calculated effective Q-factor reduces with power during this part of the transition to the superconducting regime. Below -15 dBm the superconductivity starts to dominate and the normal-state regions diminish. The Q-factor now starts to grow until the resonator surface becomes completely superconducting at approximately -40 dBm. After this, the reduction in power does not affect the cavity resonance anymore and a Q-factor order of 10$^5$ is achieved nearly independent on power.

An interesting feature such a 3D cavity can offer is the possibility to operate in a very low loss regime with a small amount of photons. In this sense, the number of photons $n_{photon}$ inside a cavity can be calculated by dividing the total electromagnetic energy into the cavity $E_{cav}$ by the energy of a single photon, as \cite{hartnett2011}:

\begin{equation}
n_{photon} = \frac{E_{cav}}{\hbar\omega}, 
\label{Eq2}
\end{equation}

\noindent where $\omega = 2\pi f$ and the total energy stored in the tunable cavity is given by

\begin{equation}
E_{cav} = P_{input}\frac{Q_{cav}}{\omega} \frac{4\beta_1}{(1 + \beta_1 + \beta_2)^2}.
\label{Eq3}
\end{equation}

In the Eq. (\ref{Eq3}) $\beta_1$ and $\beta_2$ are the coupling coefficients of the cavity input and output ports, respectively. They must be considered because the calculation of the energy into the cavity has to account for the strength of its coupling to the external circuit, as well as the loading of the Q-factor by it. At 6.1 GHz the couplings were  $\beta_1 \approx \beta_2 \approx 0.1$ and Fig. \ref{fig3} (c) presents how the photon number would vary for different input power.

Measurements were undertaken at 10 mK with an input power of -127 dBm; such conditions combined with a Q-factor of $10^5$ implicates that we achieved approximately 35 microwave photons on average stored within the cavity, just 15 dB from the single photon level. This level of photons has been shown to be sufficient for cavity characterization for quantum applications \cite{creedon2011, kostylev2016}.

\section{Conclusions}

In this work we presented a finely tunable microwave three-dimensional cavity based on a reentrant design, which was electronically controlled over a range of several hundred of megahertz. This system demonstrated effective performance at dilution refrigerator temperatures and capability of operation in the superconducting regime, leading to the achievement of high-Q microwave modes. The resonator also presented thermal nonlinear effects associated to the breakdown of the superconductivity due high power microwave driving. In contrast, its low power operation illustrated the high-Q operation with few photons in the cavity. A superconducting fine-tunable 3D microwave cavity with such characteristics is a desirable device for quantum-enabled experiments and can offer practical contribution to the research area. 

\begin{acknowledgments}

This research is supported by the Australian Research Council grant no. CE110001013 and by the Conselho Nacional de Desenvolvimento Cient\'ifico e Tecnol\'ogico (CNPq –  Brazil).

\end{acknowledgments}



%

\end{document}